\documentclass[final, 3p, times]{elsarticle}
\usepackage{amsmath,amssymb,booktabs,graphicx,xcolor}
\usepackage{placeins}
\usepackage[colorlinks=true]{hyperref}

\newcommand{\tableformat}{\centering\small}

\makeatletter
\newcommand{\appendixlabel}[1]{%
  \begingroup
  \protected@edef\@currentlabel{\@Alph\c@section}%
  \label{#1}%
  \endgroup
}
\makeatother

\begin{document}
\begin{frontmatter}
  \title{A Rapid Integrated Tokamak Modelling Approach for Coupled Profile and Equilibrium Evolution}
  \author[dlut,veloalpha]{Ruohan Zhang}
  \ead{zhangtakeda@gmail.com}
  \author[veloalpha]{Huasheng Xie\corref{cor1}}
  \ead{huashengxie@gmail.com}
  \author[dlut]{Feng Wang}
  \author[dlut]{Zheng-Xiong Wang\corref{cor1}}
  \ead{zxwang@dlut.edu.cn}
  \cortext[cor1]{Corresponding authors.}
  \address[dlut]{Key Laboratory of Materials Modification by Beams of the Ministry of Education, School of Physics, Dalian University of Technology, Dalian 116024, China}
  \address[veloalpha]{Beijing VeloAlpha Technology Co., Ltd., Beijing 100080, China}
  \begin{abstract}
    During plasma heating and current ramps the thermal and magnetic states of a tokamak evolve on different timescales. We present a rapid integrated tokamak modelling approach, which couples a variational fixed-boundary Grad--Shafranov equilibrium to sparse particle and energy flux matching and to weak-form current diffusion. An ITER-like pure-deuterium scenario is examined: a current ramp from 10 to 15~MA over 20~s, a 20~s flat-top and a 20~s ramp-down to 14~MA, under 80~MW of electron-cyclotron heating applied at the start. The toroidal current density at the pedestal top more than doubles during the ramp-up and then decreases during the flat-top as current redistributes inward. Thermal stored energy rises from 185.52 to 252.70~MJ and approaches its final value within the first few seconds. An independent calculation with the Fusion Synthesis Engine (FUSE) gives a stored-energy difference of 0.02\% at 60~s, whereas the largest profile differences occur in the core current density. Halving the coupling interval changes stored energy by at most 0.006\% at the selected radial resolution. The 60~s pulse, advanced through 600 coupling intervals of 0.1~s, takes 8.62~s on an Apple M4 Pro, which is at least two orders of magnitude faster than conventional integrated tokamak modelling solvers.
  \end{abstract}
\end{frontmatter}

\section{Introduction}
\label{sec:introduction}
Heating and current ramps couple the thermal and magnetic evolution of tokamak plasmas. Temperature and density determine pressure, while temperature, collisionality and magnetic geometry affect resistivity and bootstrap current. Current redistribution changes the safety factor, magnetic shear and equilibrium geometry entering the transport fluxes. Integrated modelling captures these feedbacks by advancing transport, current diffusion and the magnetic equilibrium together. Bourdelle\cite{bourdelle2025} reviews progress in validating such integrated calculations and the remaining challenges for ITER operation and reactor design, including pedestal transport and the feedback between core profiles and fusion heating.

Existing integrated models cover this coupling at different points on the fidelity--cost trade-off. CRONOS\cite{artaud2010cronos} combines one-dimensional transport with two-dimensional equilibrium and source models. ASTRA coupled to SPIDER\cite{fable2013astra} has been used to study free-boundary evolution during ASDEX Upgrade current ramps, and STEP\cite{lyons2023step} couples transport, equilibrium, pedestal and stability calculations. TRANSP\cite{pankin2025transp,pankin2026predictive} supports interpretive and predictive transport, current diffusion and heating calculations. The Integrated Modelling \& Analysis Suite (IMAS)\cite{imbeaux2015imas} provides a common description of the physical data exchanged in such calculations. At higher transport fidelity, TRINITY\cite{barnes2010trinity} and GENE--Tango\cite{disiena2022genetango} couple profile evolution to local and global gyrokinetic turbulence, respectively.

Reduced transport models target rapid scenario development and plasma control. FASTRAN\cite{fastran2017}, RAPTOR\cite{felici2011raptor,felici2012raptor} and TORAX\cite{citrin2024torax} provide fast profile evolution, with RAPTOR also aimed at real-time reconstruction and discharge optimisation. The Fusion Synthesis Engine (FUSE)\cite{meneghini2024fuse} combines models of different fidelity for plasma and fusion-plant studies. METIS\cite{artaud2018metis} combines simplified transport normalised to confinement scaling laws with current diffusion and a three-moment equilibrium description, for minute computation time scenario simulation. These approaches motivate rapid coupled evolution with a systematically refinable description of magnetic geometry.

In this work the VTS (Veloce Tokamak Simulator) is applied to thermal relaxation and current redistribution during a heated current pulse, with the equilibrium geometry responding to the evolving pressure and current. The magnetic geometry uses the Miller-extended-harmonic (MXH) Chebyshev representation\cite{xie2026mxh} together with the VEQ variational Grad--Shafranov formulation\cite{zhang2026veq}; particle and energy flux matching and current diffusion are coupled to that geometry. The reduced physical model comprises mixed Bohm/gyro-Bohm transport\cite{erba1998}, Chang--Hinton ion heat transport\cite{chang1982} and analytical neoclassical current closures\cite{redl2021}. A separately converged FUSE calculation, run with matched sources, pedestal profiles and current schedule, provides an independent comparison. Time refinement and particle and energy balances establish the numerical sensitivity of the predicted evolution. This work demonstrates that the coupled profile and equilibrium evolution of a typical integrated tokamak modelling case can be simulated in a matter of seconds by the proposed rapid approach.

\section{Physical model and numerical formulation}
\label{sec:model}

VTS couples an axisymmetric magnetic equilibrium, particle and energy transport, and current diffusion within a prescribed plasma boundary. The evolving profiles are the ion densities, a common ion temperature, the electron temperature and the safety factor. Pressure and current determine the magnetic equilibrium, whose geometry feeds back into the transport and current-diffusion equations. For the present comparison the physical closures follow those used in FUSE\cite{meneghini2024fuse}.

Transport and current profiles are expressed in the normalized toroidal-flux coordinate $\rho=(\Phi/\Phi_b)^{1/2}$, with $\Phi=0$ on the magnetic axis and $\Phi_b$ its value at the last closed flux surface. Let $V(\rho)$ and $S(\rho)$ denote the enclosed volume and poloidal cross-sectional area, and $A(\rho)$ the flux-surface area; a subscript $\rho$ denotes radial differentiation. Temperatures are expressed in eV, so that the pressure is $P=e(n_eT_e+n_{\mathrm{ion}}T_i)$, where $e$ is the elementary charge and $n_{\mathrm{ion}}=\sum_s n_s$. Quasineutrality gives $n_e=\sum_s Z_s n_s$, with the species charges $Z_s$ prescribed.

\subsection{Fixed-boundary magnetic equilibrium}
\label{sec:equilibrium}

The magnetic geometry is obtained from the MXH--Chebyshev flux-surface representation\cite{xie2026mxh} together with the VEQ variational Grad--Shafranov formulation\cite{zhang2026veq}. Its shape coefficients adjust to the evolving pressure and current profiles while preserving the prescribed boundary.

The poloidal flux per radian $\psi$ satisfies the Grad--Shafranov equation
\begin{equation}
  \Delta^*\psi=-\mu_0R^2P_\psi-FF_\psi,
  \qquad
  \Delta^*=R\partial_R(R^{-1}\partial_R)+\partial_Z^2,
  \label{eq:gs}
\end{equation}
where $F=RB_\phi$ and $P$ is the pressure. We follow the VEQ notation, writing its radial surface label as $r\in[0,1]$ to distinguish it from the normalized toroidal-flux coordinate $\rho$. The flux surfaces are represented by
\begin{equation}
  \begin{aligned}
    R            & =R_0+a[h(r)+r\cos\bar{\theta}(r,\theta)],                 \\
    Z            & =Z_0+a[v(r)-r\kappa(r)\sin\theta],                        \\
    \bar{\theta} & =\theta+c_0(r)+\sum_{m=1}^{M}
                                  [c_m(r)\cos(m\theta)+s_m(r)\sin(m\theta)].
  \end{aligned}
  \label{eq:surfaces}
\end{equation}
Here $(R_0,Z_0)$ specifies the reference centre and $a$ is the boundary minor-radius scale. The shaping profiles are the elongation parameter $\kappa(r)$ and triangularity parameter $\delta(r)=\sin s_1(r)$. The radial shape functions use Chebyshev expansions with prescribed edge values and mode-dependent axis regularity. Their interior variations contain the factor $(1-r^2)$, which preserves the boundary.

The equilibrium closure prescribes $P(\rho)$ and the toroidal-current-density profile $j_{\mathrm{tor}}(\rho)$, evaluated at $\rho(r)$ for each trial geometry. To express the closure in physical units, define
\begin{equation}
  \begin{aligned}
    J                & =R_\theta Z_r-R_rZ_\theta,                                                 &
    g_{\theta\theta} & =R_\theta^2+Z_\theta^2,                                                                                                                  \\
    \hat{K}          & =\frac{1}{2\pi}\int_0^{2\pi}\frac{g_{\theta\theta}}{JR}\,\mathrm{d}\theta,
                     & \hat{L}_r                                                                  & =\frac{1}{2\pi}\int_0^{2\pi}\frac{J}{R}\,\mathrm{d}\theta .
  \end{aligned}
  \label{eq:geometry}
\end{equation}
With $S_r=\int J\,\mathrm{d}\theta$ and $V_r=2\pi\int RJ\,\mathrm{d}\theta$, the current closure becomes
\begin{equation}
  \begin{aligned}
    I_{\mathrm{tor}}(r) & =\int_0^r j_{\mathrm{tor}}(\xi)S_r(\xi)\,\mathrm{d}\xi,
                        & \psi_r                                                  & =\frac{\mu_0I_{\mathrm{tor}}}{2\pi\hat{K}}, \\
    FF_\psi             & =-\frac{\mu_0}{\hat{L}_r}
    \left(\frac{j_{\mathrm{tor}}S_r}{2\pi}
    +\frac{V_rP_\psi}{4\pi^2}\right),
                        & P_\psi                                                  & =\frac{P_r}{\psi_r}.
  \end{aligned}
  \label{eq:current-closure}
\end{equation}
The imposed total current fixes $I_{\mathrm{tor}}(1)=I_p$, and regular continuation supplies the magnetic-axis limits. The diagnostic current densities and safety factor are
\begin{equation}
  j_{\mathrm{tor}}=\frac{\mathrm{d}I_{\mathrm{tor}}}{\mathrm{d}S},
  \qquad j_{\mathrm{total}}=\frac{\langle\boldsymbol{j}\cdot\boldsymbol{B}\rangle}{B_0},
  \qquad q=\frac{1}{2\pi}\frac{\mathrm{d}\Phi}{\mathrm{d}\psi},
  \label{eq:equilibrium-diagnostics}
\end{equation}
where $B_0$ is the signed reference toroidal field and $\boldsymbol{j}$ the local current density. For a scalar $f$, the flux-surface average in the chosen coordinates is $\langle f\rangle=\int_0^{2\pi}fRJ\,\mathrm{d}\theta/\int_0^{2\pi}RJ\,\mathrm{d}\theta$. The flux derivatives are $P_\psi=\mathrm{d}P/\mathrm{d}\psi$ and $FF_\psi=F\,\mathrm{d}F/\mathrm{d}\psi$.

VEQ determines the shape coefficients by projecting the Grad--Shafranov residual onto flux-surface displacements induced by each coefficient\cite{zhang2026veq}. Legendre radial and uniform poloidal quadrature, with spectral differentiation, evaluate these variational equations; the Powell hybrid method solves the resulting system. This closes the feedback from pressure and current to magnetic geometry, including the metrics entering transport and current diffusion. Parameterized variational equilibria have also been used in SuperCode for rapid tokamak systems analysis\cite{haney1992supercode}.

\subsection{Particle and energy transport}
\label{sec:transport}

For each species or thermal channel $c$, define the local inventory density $U_c$, source density $s_c$, and outward flux $f_c$. The particle channels have $U_s=n_s$ and $f_s=\Gamma_s$, while the thermal channels have
\begin{equation}
  U_i=\tfrac32 e n_{\mathrm{ion}}T_i,\qquad
  U_e=\tfrac32 e n_eT_e,\qquad f_i=Q_i,\quad f_e=Q_e.
\end{equation}
During each transport advance the equilibrium geometry is fixed, and conservation of particles and thermal energy gives
\begin{equation}
  \left.\frac{\partial}{\partial t}\right|_\rho
  \int_0^\rho U_c(\rho',t)V_\rho(\rho',t)\,\mathrm{d}\rho'
  =\int_0^\rho s_cV_\rho\,\mathrm{d}\rho'-A(\rho)f_c(\rho).
  \label{eq:conservation}
\end{equation}
Here $f_c$ is the outward flux through a stationary magnetic surface. The vanishing enclosed volume and surface area enforce zero throughput at the axis. Density and temperature profiles at and beyond the pedestal top, $\rho_{\mathrm{ped}}$, are prescribed.

Following the flux-matching formulation used in TGYRO\cite{candy2009tgyro} and FUSE, we solve for the logarithmic gradients $z_c=\partial_\rho\ln y_c$ of $y_c\in\{n_s,T_i,T_e\}$ on sparse flux surfaces. Integrating inward from the pedestal gives
\begin{equation}
  y_c(\rho)=y_c(\rho_{\mathrm{ped}})
  \exp\left[-\int_\rho^{\rho_{\mathrm{ped}}}z_c(\xi)\,\mathrm{d}\xi\right],
  \label{eq:reconstruction}
\end{equation}
and a cumulative-volume operator $C$ integrates the reconstructed profiles and sources on a denser radial grid. The steady equations
\begin{equation}
  \mathcal{R}_{c,j}(\boldsymbol{z})=
  [C\boldsymbol{s}_c(\boldsymbol{z})]_j
  -A_j f_{c,j}(\boldsymbol{z})=0
  \quad\text{for all }c,j
  \label{eq:transport-system}
\end{equation}
therefore balance the integrated source inside each flux surface against the transport through it. Ohmic heating, radiation and electron--ion energy exchange are recomputed from each trial profile when enabled, whereas the external sources of Sec.~\ref{sec:closures} are held fixed during an advance.

The particle and energy balances, including the inventory changes in time-dependent stages, are solved with a structured Newton--Krylov method; \ref{app:newton-krylov} summarizes the method and its convergence tests. Time-dependent transport advances the cumulative inventories of Eq.~\eqref{eq:conservation} with the trapezoidal rule combined with the second-order backward differentiation formula (TR-BDF2)\cite{bank1985,hosea1996}. Its implicit stages accommodate stiff relaxation, and its embedded error estimate refines the step during rapid profile changes; \ref{app:time-integration} describes the adaptive step-size strategy. Electron--ion exchange cancels in the total thermal-energy balance. The geometry is updated between transport advances.

\subsection{Transport closures and external sources}
\label{sec:closures}

The transport closure combines a mixed Bohm/gyro-Bohm (BGB) turbulent model\cite{erba1998} with Chang--Hinton ion heat transport\cite{chang1982}. With $g_y=-\partial_\rho\ln y$, $g_{p,e}=g_{n_e}+g_{T_e}$ and $g_{p,i}=g_{n_{\mathrm{ion}}}+g_{T_i}$, the BGB particle and thermal fluxes are
\begin{equation}
  \begin{aligned}
    \Gamma_s & =n_s(Dg_{n_s}-v_{\mathrm{in}}),
             & \Gamma_e                                        & =\sum_s Z_s\Gamma_s,      \\
    Q_i      & =\chi_i n_{\mathrm{ion}}eT_i g_{p,i},
             & Q_e                                             & =\chi_e n_e eT_e g_{p,e}, \\
    D        & =(1-0.7\rho)\frac{\chi_e\chi_i}{\chi_e+\chi_i},
             & v_{\mathrm{in}}                                 & =\frac{DA^2}{2VV_\rho}.
  \end{aligned}
  \label{eq:bgb-flux}
\end{equation}
These coefficients multiply gradients taken with respect to the dimensionless coordinate $\rho$. All species share the diffusion and inward-pinch coefficients, while each particle flux responds to its own density gradient. In the radial-flux convention of Eq.~\eqref{eq:bgb-flux}, the BGB diffusivities are
\begin{equation}
  \begin{aligned}
    \chi_{e,B} & =2\times10^{-4}\frac{a_{\mathrm{cm}}q^2}{|B_0|}T_e g_{p,e},
               & \chi_{e,gB}                                                 & =5\times10^{-6}\frac{T_e^{3/2}}{|B_0|}g_{T_e},                         \\
    \chi_{i,B} & =2\chi_{e,B},                                               & \chi_{i,gB}                                    & =\tfrac12\chi_{e,gB}, \\
    \chi_e     & =m(c_{eB}\chi_{e,B}+c_{egB}\chi_{e,gB}),
               & \chi_i                                                      & =m(c_{iB}\chi_{i,B}+c_{igB}\chi_{i,gB}),
  \end{aligned}
\end{equation}
with $T_e$ in eV, $B_0$ in tesla and $a_{\mathrm{cm}}$ in centimetres.

The present study uses $(m,c_{eB},c_{egB},c_{iB},c_{igB})= (0.01,0.01,50,0.001,1)$, which fix an idealised transport scenario rather than a validated confinement scaling, and the gradients are evaluated on the sparse transport surfaces.

Ohmic heating, radiation and electron--ion exchange are reevaluated at each trial profile. The radiative loss comprises bremsstrahlung, line radiation and synchrotron emission; line cooling uses the tabulated fits supplied by IMAS.jl, and the synchrotron model assumes a wall reflection coefficient of 0.8.

\subsection{Current diffusion and non-inductive current}
\label{sec:current}

Current diffusion is formulated in terms of the rotational transform $\iota=1/q$. Let $K=\langle|\nabla\rho|^2/R^2\rangle$ and $g_1=\langle R^{-2}\rangle$, where brackets denote a flux-surface average. At the fixed geometry used during each diffusion advance, the flux-surface-averaged induction equation\cite{jardin2010} takes the weak form
\begin{equation}
  \frac{\mathrm{d}}{\mathrm{d}t}\int_0^1\nu\Phi_\rho\iota\,\mathrm{d}\rho
  =-\int_0^1\nu_\rho\mathcal E\,\mathrm{d}\rho
  +[\nu\mathcal E]_0^1,
  \qquad \mathcal E=a_b\iota_\rho+a_d\iota-v_{\mathrm{ni}},
  \label{eq:current-weak}
\end{equation}
where $\nu$ is a time-independent test function and $\mathcal E$ the loop voltage, with
\begin{equation}
  \begin{aligned}
    a_b             & =\frac{\eta\Phi_\rho K}{\mu_0g_1},                            \\
    a_d             & =\frac{\eta}{\mu_0g_1}
    \left[(\partial_\rho\Phi_\rho)K+\Phi_\rho K_\rho
      +\Phi_\rho K\left(\frac{\partial_\rho V_\rho}{V_\rho}
      -\frac{F_\rho}{F}\right)\right],                          \\
    v_{\mathrm{ni}} & =\frac{2\pi\eta B_0(j_{\mathrm{bs}}+j_{\mathrm{dr}})}{Fg_1} .
  \end{aligned}
  \label{eq:current-coefficients}
\end{equation}
Here $\eta$ is the parallel resistivity, and $j_{\mathrm{bs}}$ and $j_{\mathrm{dr}}$ are the equivalent parallel bootstrap and driven current densities in the convention $\langle\boldsymbol{j}\cdot\boldsymbol{B}\rangle=B_0j$. The plotted ohmic component is $j_{\mathrm{ohm}}=j_{\mathrm{total}}-j_{\mathrm{bs}}-j_{\mathrm{dr}}$. Resistivity and bootstrap current follow the analytical neoclassical fits of Redl et al.\cite{redl2021}, with implementation guidance from IMAS.jl. Where $|q|<1$, the diffusion operator uses an effective resistivity held at its value at the outermost violating radial node throughout the enclosed region. The correction affects only the diffusion operator, so the underlying neoclassical resistivity is retained for the physical diagnostics.

Following the QED.jl discretization, cubic Hermite interpolation supplies the nodal values and derivatives of $\iota$, so that with basis functions $H_j$ the coefficients $\boldsymbol{w}$ satisfy
\begin{equation}
  \frac{\mathrm{d}}{\mathrm{d}t}(M\boldsymbol{w})
  =Y\boldsymbol{w}+\boldsymbol{b},
  \qquad M_{ij}=\int_0^1\Phi_\rho H_iH_j\,\mathrm{d}\rho,
\end{equation}
where $Y$ and $\boldsymbol b$ are the diffusion matrix and non-inductive forcing obtained from Eq.~\eqref{eq:current-weak}. Axis regularity and the prescribed total current impose
\begin{equation}
  \iota_\rho(0)=0,\qquad
  \iota(1)=\left.\frac{4\pi^2\mu_0I_p}{\Phi_\rho V_\rho K}\right|_{\rho=1}.
  \label{eq:current-boundary}
\end{equation}
At steady state, $-Y\boldsymbol w=\boldsymbol b$ is solved together with Eq.~\eqref{eq:current-boundary}. Time-dependent current diffusion also uses TR-BDF2, with its error estimate and adaptive step-size strategy described in \ref{app:time-integration}. During each time-dependent advance the geometry and the neoclassical coefficients are held fixed, so the imposed $I_p(t)$ alone drives the boundary condition. Resistivity and bootstrap current are refreshed at the next coupling step.

\subsection{Coupled solution}
\label{sec:coupling}

The steady calculation iterates particle and energy balance, current diffusion and magnetic equilibrium. Updated densities and temperatures set the pressure, resistivity and bootstrap current, and the resulting pressure and current determine the next geometry. The iteration is of Picard type with a relaxation factor of 0.5, and convergence is monitored on the pressure, electron density, safety factor and poloidal flux:
\begin{equation}
  \epsilon_{\mathrm{coupled}}=
  \max_{f\in\{P,n_e,q,\psi\}}
  \frac{\|f^{k+1}-f^k\|_2}{\max(\|f^{k+1}\|_2,\|f^k\|_2)}
  \leq10^{-6}.
  \label{eq:coupled-tolerance}
\end{equation}
The 60~s pulse is divided into 600 coupling intervals of 0.1~s. Within each interval, current diffusion advances, the equilibrium is updated, particle and heat transport advance, and the equilibrium is updated once more; adaptive substeps remain confined to the interval. Current and thermal profiles are advanced from the interval-start state to a common endpoint, and both codes follow this sequence. The geometry is held fixed within each solve and changes only at the equilibrium updates. VTS varies the imposed current linearly across an interval, whereas FUSE applies its endpoint value throughout the diffusion solve. Each VTS equilibrium solve reuses its own preceding converged solution, and the transport solve reuses the preceding converged gradients and implicit-stage solutions.

\section{Model scenario and comparison procedure}
\label{sec:calculation}

\subsection{Heating and current programme}

The time-dependent calculation starts from a self-consistent steady state at $I_p=10$~MA. The plasma current then increases linearly to 15~MA over 20~s, remains constant for a further 20~s, and decreases to 14~MA over the final 20~s:
\begin{equation}
  I_p(t)=
  \begin{cases}
    10+0.25t,      & 0\leq t\leq20~\mathrm{s}, \\
    15,            & 20<t\leq40~\mathrm{s},    \\
    15-0.05(t-40), & 40<t\leq60~\mathrm{s},
  \end{cases}
  \quad [I_p]=\mathrm{MA},\quad [t]=\mathrm{s}.
  \label{eq:current-waveform}
\end{equation}
The thermal and current profiles evolve continuously through both phase transitions; during the flat-top only the total current is held constant.

\subsection{Matched inputs and current transfer}

The scenario is the pure-deuterium ITER model case distributed with FUSE~1.1.5, and the two codes are run independently with matched external sources, prescribed pedestal profiles and current schedule. The external heating comprises neutral beam injection (NBI), lower-hybrid current drive (LHCD), ion-cyclotron resonance heating (ICRH) and electron-cyclotron resonance heating (ECRH), all taken from the same IMAS dataset; ECRH is abbreviated as EC below. Each calculation retains its own equilibrium representation and discretization and starts from its separately converged 10~MA steady state, with the VTS boundary fitted to the initialized outer flux surface. Table~\ref{tab:configuration} gives the principal settings.

The FUSE calculation includes two current-consistency corrections submitted upstream:\footnote{\href{https://github.com/ProjectTorreyPines/QED.jl/pull/40}{QED.jl pull request 40}, commit \texttt{360a4cc}, and \href{https://github.com/ProjectTorreyPines/TEQUILA.jl/pull/72}{TEQUILA.jl pull request 72}, commit \texttt{794de4b}.} QED retains the native radial grid of the input toroidal-current profile, and the TEQUILA Picard equilibrium solve checks the reconstructed total current in addition to the magnetic-axis flux convergence. These corrections address input-profile projection and total-current convergence. Residual errors in the repeated IMAS--QED current transformations are examined in Sec.~\ref{sec:transient}.

\begin{table}[!htbp]
  \tableformat
  \caption{Principal settings of the heated current-pulse scenario.}
  \label{tab:configuration}
  \begin{tabular}{@{}lp{0.62\textwidth}@{}}
    \toprule
    Quantity                   & Value                                                                                                                                                \\
    \midrule
    Boundary shape             & $R_0=6.1918$~m, $a=2.0037$~m, boundary elongation $\kappa_b=1.8283$                                                                                  \\
    Reference field            & $B_0=-5.3002$~T in COCOS~1\cite{sauter2013cocos}                                                                                                     \\
    Current schedule           & 10~MA initial state; ramp to 15~MA at 20~s; flat-top to 40~s; ramp to 14~MA at 60~s                                                                  \\
    Thermal species            & Deuterium, $n_e=n_D$                                                                                                                                 \\
    Pedestal                   & Profiles fixed for $\rho\geq0.85$; $n_e=7.1151\times10^{19}$~m$^{-3}$ and $T_e=T_i=2.4164$~keV at $\rho=0.85$; $T_e=T_i=0.300$~keV at the separatrix \\
    External heating           & 80~MW central EC deposition from $t=0^+$; other source densities fixed                                                                               \\
    Transport surfaces         & 17 uniformly spaced surfaces on $0.05\leq\rho\leq0.85$                                                                                               \\
    Equilibrium discretization & 70 shape coefficients; $32\times32$ quadrature                                                                                                       \\
    Transport model            & Mixed Bohm/gyro-Bohm turbulence and Chang--Hinton ion heat transport                                                                                 \\
    Current model              & Redl et al.\cite{redl2021} resistivity and bootstrap current; $|q|=1$ trigger for the resistivity correction                                         \\
    Nonlinear solvers          & Powell hybrid (equilibrium) and Newton--Krylov (transport)                                                                                           \\
    Time integration           & Adaptive TR-BDF2 in both subsystems; relative and absolute time tolerances $10^{-4}$ and $10^{-8}$                                                   \\
    Coupling interval          & 0.1~s, with synchronized current and thermal endpoints                                                                                               \\
    \bottomrule
  \end{tabular}
\end{table}

The prescribed edge-temperature profile retains the original pedestal-top value at $\rho=0.85$, and for $\rho>0.85$ a quadratic correction raises the separatrix temperature from 80 to 300~eV,
\begin{equation}
  T_{e,i}^{\rm edge}(\rho)=T_{e,i}^{\rm original}(\rho)
  +220\left(\frac{\rho-0.85}{0.15}\right)^2~\mathrm{eV},
  \label{eq:edge-temperature}
\end{equation}
whose value and first derivative match the original profile at the pedestal top. The edge density profile retains its original values. Together with the ramp-down to 14~MA, this temperature profile keeps the toroidal current density positive throughout the sampled pulse.

At $t=0^+$ the initialized EC source is replaced by a Gaussian deposition profile proportional to $\exp[-(\rho-0.25)^2/(2\times0.12^2)]$ carrying 80~MW of electron heating, compared with approximately 19~MW before the transition, and prescribing an equivalent parallel-current integral of 0.514~MA. The source remains active for the whole 60~s evolution. Its shape is fixed in $\rho$: at each evaluation the power density is renormalized by the instantaneous equilibrium volume and the driven-current density by the poloidal area, so that the prescribed integrals are retained as the geometry evolves.

Neutral-beam, ion-cyclotron, lower-hybrid and particle source densities retain their imported profiles, so their volume integrals vary with the equilibrium. Ohmic heating, radiation and electron--ion energy exchange follow the evolving profiles, and the bootstrap current is updated separately in the current closure. In the initial VTS geometry the full-plasma NBI, ICRH and LHCD integrals are 32.05, 23.47 and 9.49~MW, respectively; these are geometry-dependent integrals of the imported profiles rather than source powers recalculated by the transport solver, and the 80~MW EC source is additional to them.

Profiles from the two calculations are compared at 0, 20, 40 and 60~s: the initial snapshot precedes the heating transition, and the later three mark the ends of the ramp-up, flat-top and ramp-down phases. In the VTS COCOS-1 convention $B_0<0$ while $I_p>0$, so the signed safety factor is negative, whereas the FUSE core profiles store a positive magnitude. All safety-factor profiles are therefore plotted as $|q|$ and the cross-code comparisons use that magnitude. Profiles are sampled at 501 equally spaced values of $\rho$ over $[0,1]$, and their differences are quantified by
\begin{equation}
  \delta_f=\frac{2\|f_{\rm VTS}-f_{\rm FUSE}\|_2}{\|f_{\rm VTS}\|_2+\|f_{\rm FUSE}\|_2},
  \label{eq:profile-difference}
\end{equation}
with $f=|q|$ for the safety factor; the same expression applies to the scalar stored energy. The norm weights the sampled radial nodes equally.

\section{Results}
\label{sec:results}

\subsection{Self-consistent steady state}
\label{sec:steady}

\begin{figure}[!htbp]
  \centering
  \includegraphics[width=1.00\textwidth]{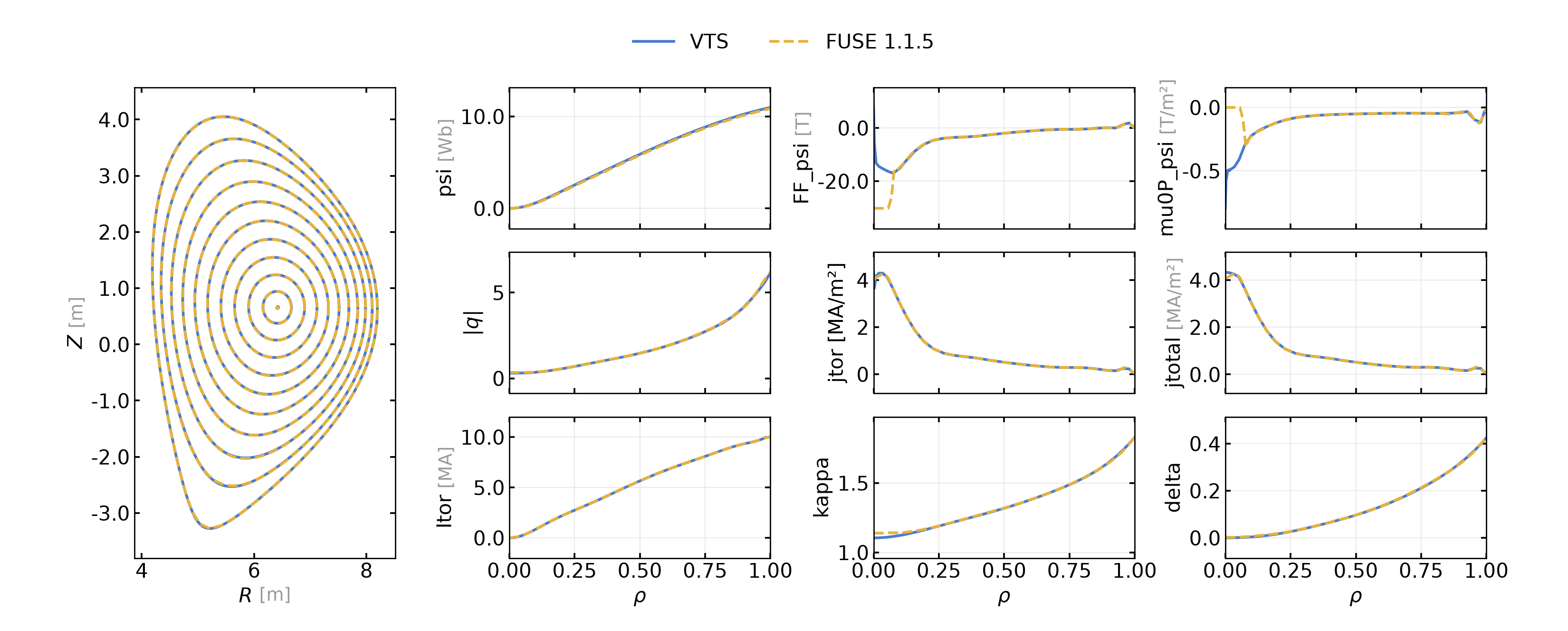}
  \caption{Independently converged 10~MA steady equilibria from VTS (blue solid) and FUSE (yellow dashed). The left panel compares the flux surfaces at equal normalized toroidal-flux radius. Against $\rho$, the remaining panels show $\psi-\psi(0)$, $FF_\psi$, $\mu_0P_\psi$, $|q|$, the toroidal and total current densities, the enclosed toroidal current, the elongation and the triangularity, with the current definitions of Eq.~\eqref{eq:equilibrium-diagnostics}. VTS uses the shaping parameters $\kappa$ and $\delta=\sin s_1$, whereas FUSE supplies the surface elongation and the mean of the upper and lower triangularity; its quantities are converted to the VTS sign convention before comparison.}
  \label{fig:equilibrium-state}
\end{figure}

\begin{figure}[!htbp]
  \centering
  \includegraphics[width=1.00\textwidth]{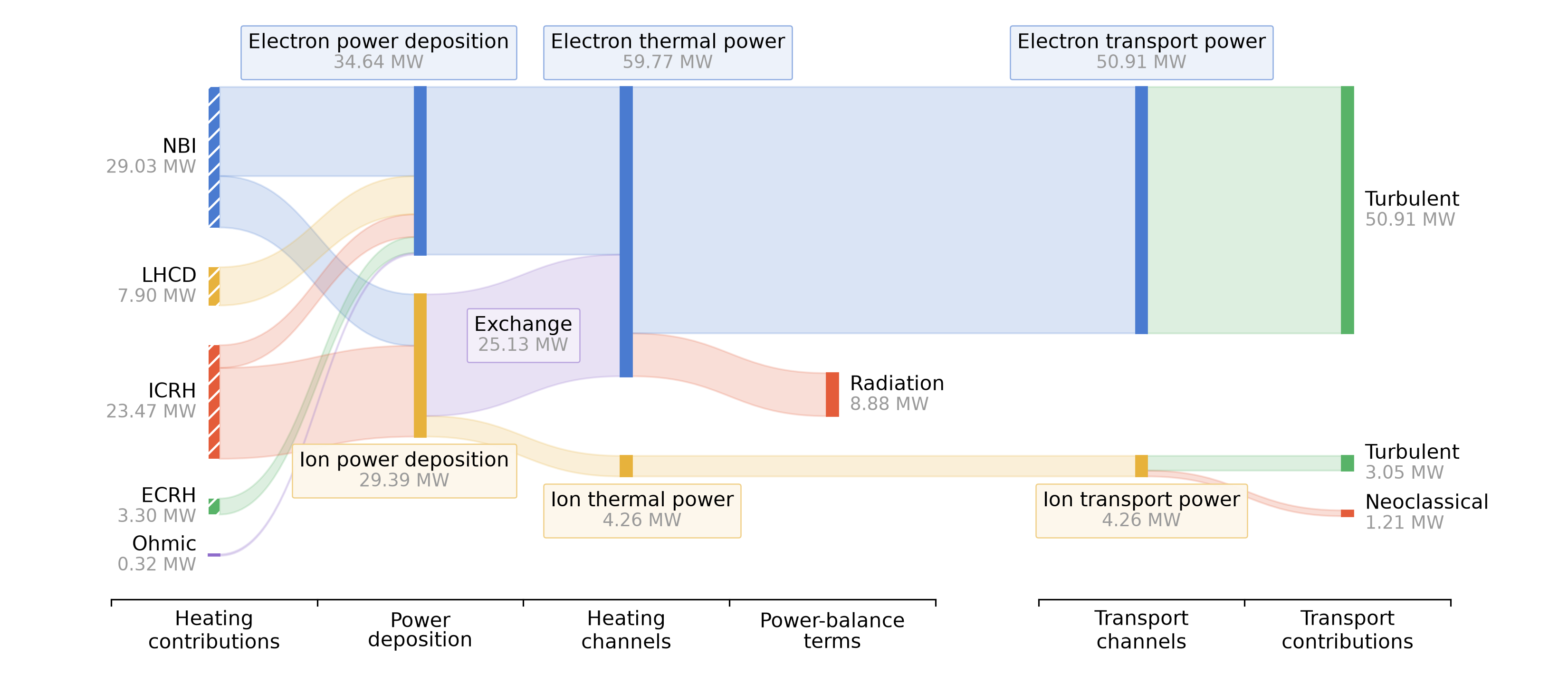}
  \caption{Converged VTS steady source--flux balance for $0\leq\rho\leq0.85$, before the heating transition. Ribbon heights represent power in MW. The four hatched heating blocks are the prescribed IMAS data-dictionary inputs of Table~\ref{tab:prescribed-power}; Turbulent denotes the mixed BGB electron and ion heat transport and Neoclassical the Chang--Hinton ion heat transport, while Radiation combines bremsstrahlung, line and synchrotron emission. Outward transport leaves the volume across $\rho=0.85$, and exchange transfers energy from ions to electrons. Contributions below 0.05~MW are omitted.}
  \label{fig:steady-power}
\end{figure}

The independently converged 10~MA states provide the initial conditions for the pulse. The central ion temperatures are 16.06~keV in VTS and 16.30~keV in FUSE, and the thermal stored energies are 185.52 and 184.74~MJ, a difference of 0.42\%. The initial electron- and ion-temperature profiles differ by 0.22\% and 0.87\% in the symmetric norm of Eq.~\eqref{eq:profile-difference}, and the central electron densities by 1.79\%. Most of the residual difference between the two initial states lies in the density profile and in the current density, which are also the quantities that dominate the transient comparison. The core-density offset is consistent with the different gradient discretizations: VTS evaluates the BGB gradients, including the geometric pinch factor $A^2/(2VV_\rho)$ of Eq.~\eqref{eq:bgb-flux}, on the sparse transport surfaces, whereas FUSE forms backward differences on its denser profile grid.

Figure~\ref{fig:equilibrium-state} compares the steady magnetic equilibria. The flux surfaces nearly coincide, with magnetic-axis major radii of 6.4135~m in VTS and 6.4096~m in FUSE, a difference of 0.06\%. The profiles of $|q|$, $j_{\mathrm{tor}}$ and the enclosed current also agree closely away from the axis. Differences remain in the central flux quantities, $\psi-\psi(0)$ and $FF_\psi$, where the two codes apply different axis treatments, and in the shaping profiles $\kappa$ and $\delta$ at large $\rho$. The VTS steady state converges in 23 coupled sweeps to a relative residual of $5.88\times10^{-7}$.

Figure~\ref{fig:steady-power} shows the converged VTS steady state and its source--flux balance inside the pedestal top, $\rho_{\rm ped}=0.85$, before the 80~MW EC transition. At steady state the enclosed thermal inventory has no time derivative, so Eq.~\eqref{eq:conservation} requires the net deposited power inside each surface to equal the outward heat flow through that surface. The interior density and temperature profiles adjust, subject to the prescribed pedestal boundary values, until this condition holds; the deuterium particle channel satisfies the analogous balance between its enclosed source and its outward flux at every surface.

The four hatched blocks denote the NBI, LHCD, ICRH and ECRH deposition profiles read directly from the IMAS data dictionary (IMASDD) and held fixed during the steady iteration. They are therefore prescribed inputs rather than heating profiles computed self-consistently within VTS, and their volume integrals depend on the geometry in which they are evaluated. For channel $c\in\{e,i\}$ and source contribution $k$,
\begin{equation}
  P_{c,k}^{\rm src}=\int_0^{\rho_{\rm ped}}s_{c,k}V_\rho\,\mathrm{d}\rho,
  \qquad P_c^{\rm out}=A(\rho_{\rm ped})Q_c(\rho_{\rm ped}),
  \label{eq:steady-power}
\end{equation}
so that the source powers are volume integrals over $0\leq\rho\leq0.85$ and the outward powers are heat fluxes integrated over the surface $\rho=0.85$. Table~\ref{tab:prescribed-power} quantifies the geometry dependence of the prescribed heating by comparing the imported cumulative powers with their integrals over the converged VTS geometry. In the diagram, Turbulent denotes the mixed BGB electron and ion heat transport, Neoclassical the Chang--Hinton ion heat transport, and Radiation the combination of bremsstrahlung, line cooling from the ADAS21 coefficient fits supplied by IMAS.jl, and synchrotron emission with a wall reflection coefficient of 0.8, all evaluated from the plasma profiles with the models of Sec.~\ref{sec:closures}.

\begin{table}[!htbp]
  \tableformat
  \caption{Prescribed external heating before the EC transition. Powers are in MW and include both electron and ion deposition, with the two NBI beam contributions combined. The IMASDD columns give the imported cumulative powers, and the VTS column integrates the same prescribed power densities over the converged geometry, corresponding to the four hatched blocks in Fig.~\ref{fig:steady-power}. The difference between the two $\rho\leq0.85$ columns reflects the volume element of the converged geometry relative to that of the import.}
  \label{tab:prescribed-power}
  \begin{tabular}{lrrr}
    \toprule
    Source & IMASDD, $\rho\leq1$ & IMASDD, $\rho\leq0.85$ & VTS, $\rho\leq0.85$ \\
    \midrule
    NBI    & 31.42               & 28.33                  & 29.03               \\
    LHCD   & 9.69                & 8.07                   & 7.90                \\
    ICRH   & 22.29               & 22.29                  & 23.47               \\
    ECRH   & 19.40               & 3.37                   & 3.30                \\
    \bottomrule
  \end{tabular}
\end{table}

The external and ohmic contributions deposit 34.64~MW in electrons and 29.39~MW in ions, of which ohmic heating supplies 0.32~MW. Collisional exchange transfers 25.13~MW from ions to electrons and electron radiation removes 8.88~MW, leaving 50.91~MW of outward electron transport and 4.26~MW in the ion channel. Exchange thus carries about 85\% of the deposited ion power to the electrons. With the chosen transport closure, most of the ion heating therefore leaves through the electron channel after collisional exchange. The thermal stored energy discussed below is $W_{\rm th}=\tfrac32\int_0^1P V_\rho\,\mathrm{d}\rho$.

\subsection{Response to sustained heating and the current pulse}
\label{sec:transient}

\begin{figure}[!htbp]
  \centering
  \includegraphics[width=1.00\textwidth]{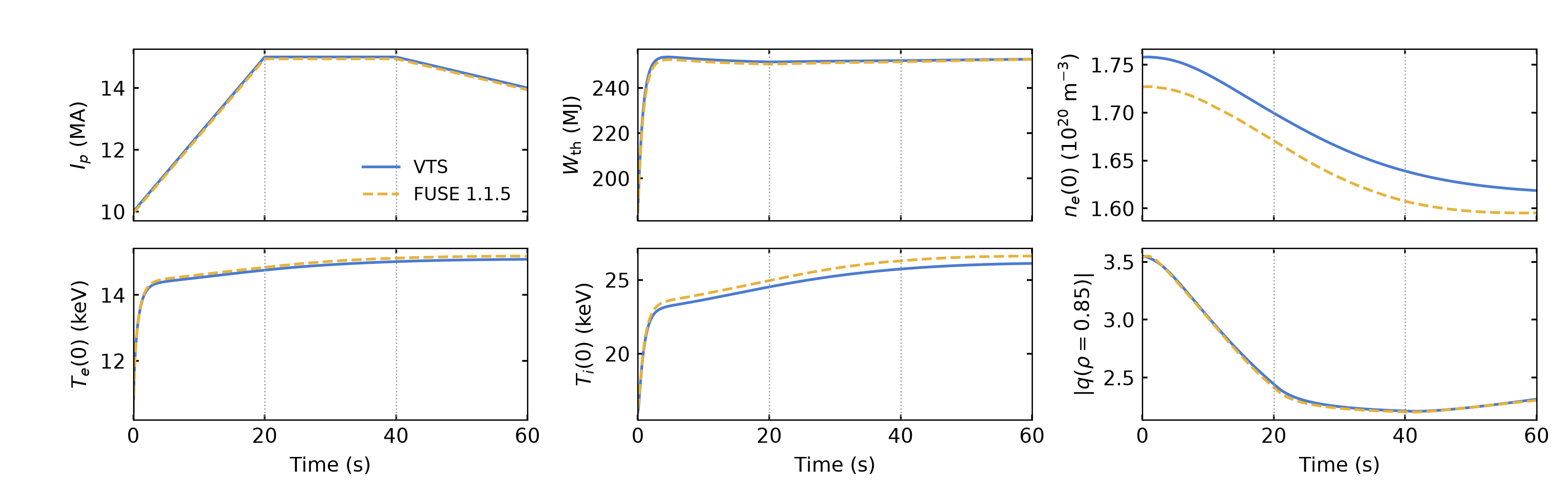}
  \caption{Evolution through the heated current pulse. The upper row shows the total plasma current, the thermal stored energy and the central electron density; the lower row shows the central electron and ion temperatures and the safety-factor magnitude at the pedestal top, $|q(0.85)|$. VTS is blue solid and FUSE is yellow dashed, and vertical dotted lines mark the phase transitions.}
  \label{fig:ramp-history}
\end{figure}

\begin{figure}[!htbp]
  \centering
  \includegraphics[width=1.00\textwidth]{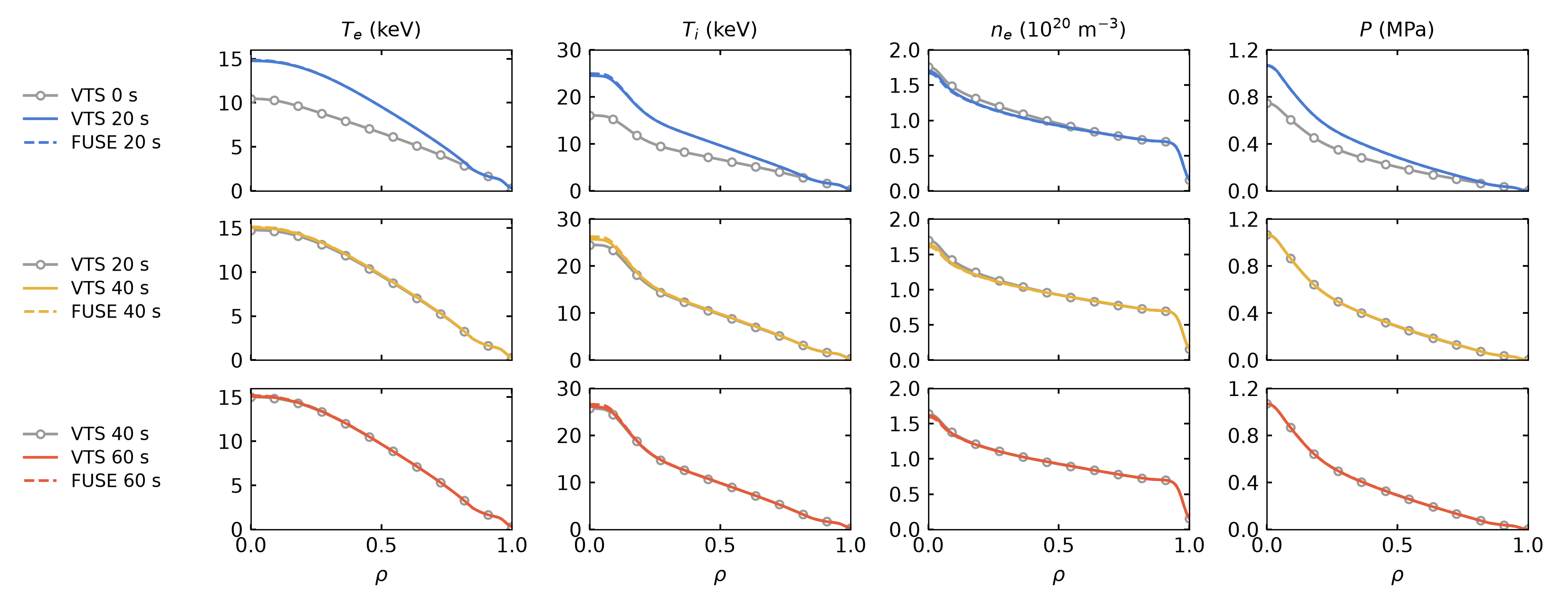}
  \caption{Thermal profiles at the phase endpoints, for VTS (solid) and FUSE (dashed). Columns show the electron temperature, ion temperature, electron density and total thermal pressure $P=e(n_eT_e+n_{\mathrm{ion}}T_i)$. Rows are the 20, 40 and 60~s endpoints in blue, yellow and red, respectively, and grey lines with open circles repeat the VTS profile at the start of each phase, i.e.\ 0, 20 and 40~s for the upper, middle and lower rows.}
  \label{fig:ramp-profiles}
\end{figure}

Figure~\ref{fig:ramp-history} follows the thermal and magnetic response to the heating transition and three-phase current pulse. Figures~\ref{fig:ramp-profiles} and \ref{fig:current-state} compare the thermal and current profiles, respectively, at the phase endpoints.

\subsubsection{Current redistribution}

During the 10--15~MA ramp the VTS toroidal current density at the pedestal top, $\rho=0.85$, rises from 0.238 to 0.634~MA\,m$^{-2}$, and the corresponding $|q(0.85)|$ falls from 3.550 to 2.443. The pedestal-top current density then drops to 0.418~MA\,m$^{-2}$ during the flat-top and $|q(0.85)|$ continues to fall to 2.211, so the edge current relaxes inward even though the total current is held constant.

The final 15--14~MA ramp-down lowers the VTS current density at $\rho=0.85$ further, to 0.275~MA\,m$^{-2}$ at 60~s, and $|q(0.85)|$ rises slightly to 2.314. Integrating the sampled $j_{\rm tor}$ over poloidal area quantifies the redistribution: the current enclosed by $\rho=0.85$ is 12.58~MA at 20~s and 13.65~MA at 40~s, so the current outside that surface falls from 2.42 to 1.35~MA during the flat-top while the total current is fixed. The concurrent fall in $|q(0.85)|$ is consistent with this increase in enclosed current; the safety factor depends on the internal current distribution and geometry, not on $I_p$ alone. During ramp-down, the enclosed current decreases by only 0.49~MA of the imposed 1~MA reduction, so about half of that reduction occurs outside $\rho=0.85$.

\begin{figure}[!htbp]
  \centering
  \includegraphics[width=1.00\textwidth]{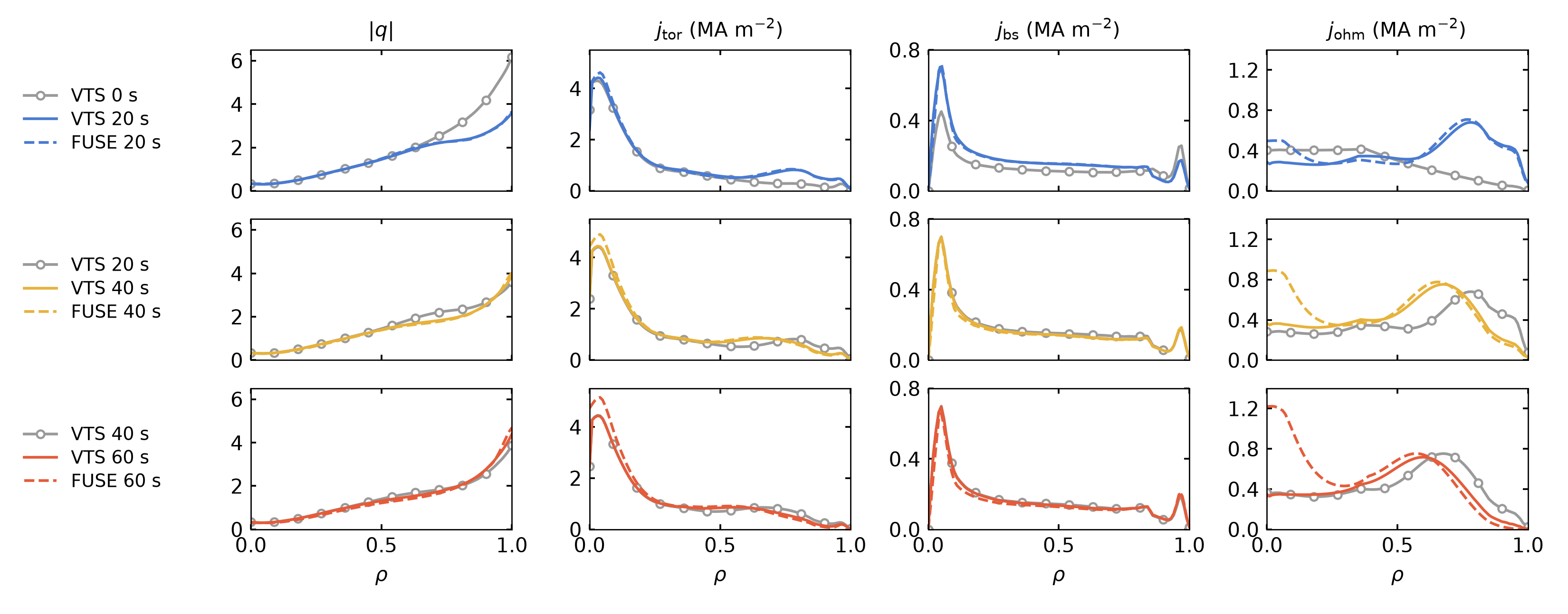}
  \caption{Current profiles over $0\leq\rho\leq1$, for VTS (solid) and FUSE (dashed), in the same layout as Fig.~\ref{fig:ramp-profiles}. Columns show the safety-factor magnitude, the toroidal current density, and the equivalent parallel bootstrap- and ohmic-current densities of Eq.~\eqref{eq:equilibrium-diagnostics}. Rows show the 20, 40 and 60~s endpoints in blue, yellow and red, and grey lines with open circles show the respective VTS phase-start profiles at 0, 20 and 40~s.}
  \label{fig:current-state}
\end{figure}

The current decomposition in Fig.~\ref{fig:current-state} shows which component changes at the pedestal top. In VTS, between 20 and 40~s the equivalent parallel ohmic-current density at $\rho=0.85$ falls from 0.531 to 0.302~MA\,m$^{-2}$, whereas the bootstrap contribution changes from 0.111 to 0.103~MA\,m$^{-2}$ and the driven current is fixed by construction. The local decrease is therefore dominated by the inductive component, as expected for current redistribution during the flat-top. These components use the equivalent parallel-current convention of Eq.~\eqref{eq:equilibrium-diagnostics}.

Both calculations retain positive sampled toroidal current density throughout the pulse, with minima of 0.0100 and 0.0144~MA\,m$^{-2}$ for VTS and FUSE. The current profiles agree well over the outer half of the plasma but diverge in the core: at 60~s they differ by 11.33\% over $0.6\leq\rho\leq1$, whereas including the core raises the difference to 16.79\% (Fig.~\ref{fig:current-state}). The corresponding $|q|$ profiles differ by 5.58\% over the outer region. The current-density difference is concentrated where the profile is least constrained by the prescribed boundary and most sensitive to the treatment of the magnetic axis.

Tests of the saved FUSE states identify a residual numerical error in the repeated transformations between the IMAS and QED current representations. Using the 40~s state with fixed equilibrium input, 100 successive round trips without current diffusion raise $j_{\mathrm{total}}(0.05)$ from 4.922 to 5.127~MA\,m$^{-2}$. This drift shows how small transfer errors can accumulate and enhance the core ohmic-current component. The two corrections described in Sec.~\ref{sec:calculation} address specific projection and convergence errors, but do not eliminate this broader inconsistency. Its contribution to the full coupled-pulse difference has not been isolated.

Differences in radial discretization and magnetic-axis treatment may also contribute. The total current reconstructed by the FUSE equilibrium solver differs from the prescribed waveform by at most 0.49\%, while the diffusion solve satisfies its own current constraint (Fig.~\ref{fig:ramp-history}).

\subsubsection{Thermal response}

In VTS, the stored energy rises from 185.52~MJ to 251.49~MJ during the first 20~s, then reaches 252.03~MJ at 40~s and 252.70~MJ at 60~s. The central electron and ion temperatures at the final time are 15.08 and 26.12~keV, compared with 10.45 and 16.06~keV initially. The central density decreases from $1.758\times10^{20}$ to $1.619\times10^{20}$~m$^{-3}$. FUSE gives 250.54, 251.50 and 252.65~MJ at the same phase endpoints. At 60~s, the electron- and ion-temperature profiles differ by 0.37\% and 1.23\%, and stored energy differs by 0.02\%.

The heating response is concentrated near the start of the pulse: $W_{\rm th}$ reaches 235.15~MJ at 1~s, completing 74\% of its net increase to 60~s, and its sampled maximum of 253.67~MJ occurs at 4.5~s, well before the current reaches 15~MA. Thermal energy adjusts much faster than the current profile redistributes; the magnetic configuration continues to evolve through the flat-top after the thermal profiles have become nearly stationary. Between 20 and 40~s the central ion temperature rises by 4.93\% while the central density falls by 3.56\%, so the central ion pressure changes by only 1.20\%. The density and temperature changes nearly cancel in their contribution to the core ion pressure.

\begin{figure}[!htbp]
  \centering
  \includegraphics[width=1.00\textwidth]{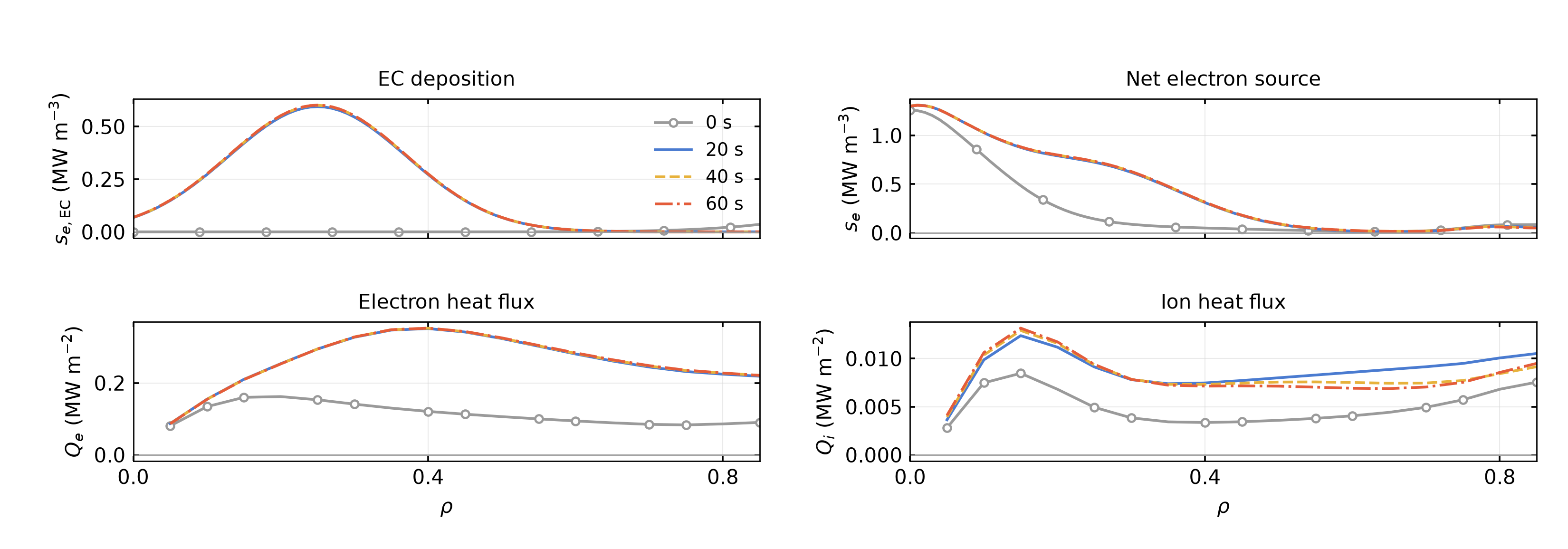}
  \caption{VTS source and outward heat-flux profiles over $0\leq\rho\leq0.85$. The initial state uses grey circles, and the ramp-up, flat-top and ramp-down endpoints use blue solid, yellow dashed and red dash-dotted curves. The upper row shows the EC electron deposition and the net electron-energy source density; the lower row shows the electron and ion heat fluxes. Source densities are per unit plasma volume and heat fluxes per unit flux-surface area.}
  \label{fig:source-flux}
\end{figure}

Figure~\ref{fig:source-flux} places this response in the power balance. Inside $\rho=0.85$ the EC contribution increases from 3.30~MW in the initial state to approximately 80~MW after the heating transition. By 5~s, outward thermal transport has increased from 55.17 to 129.93~MW, whereas radiation has increased from 8.83 to 11.23~MW. The additional heating is thus balanced mainly by increased outward heat transport rather than by radiation, allowing stored energy to approach a plateau under sustained heating. Ohmic heating contributes less than 0.4~MW at all three phase endpoints, so its variation is small against the 80~MW of injected power.

\subsection{Numerical sensitivity and energy balance}
\label{sec:verification}

\begin{figure}[!htbp]
  \centering
  \includegraphics[width=1.00\textwidth]{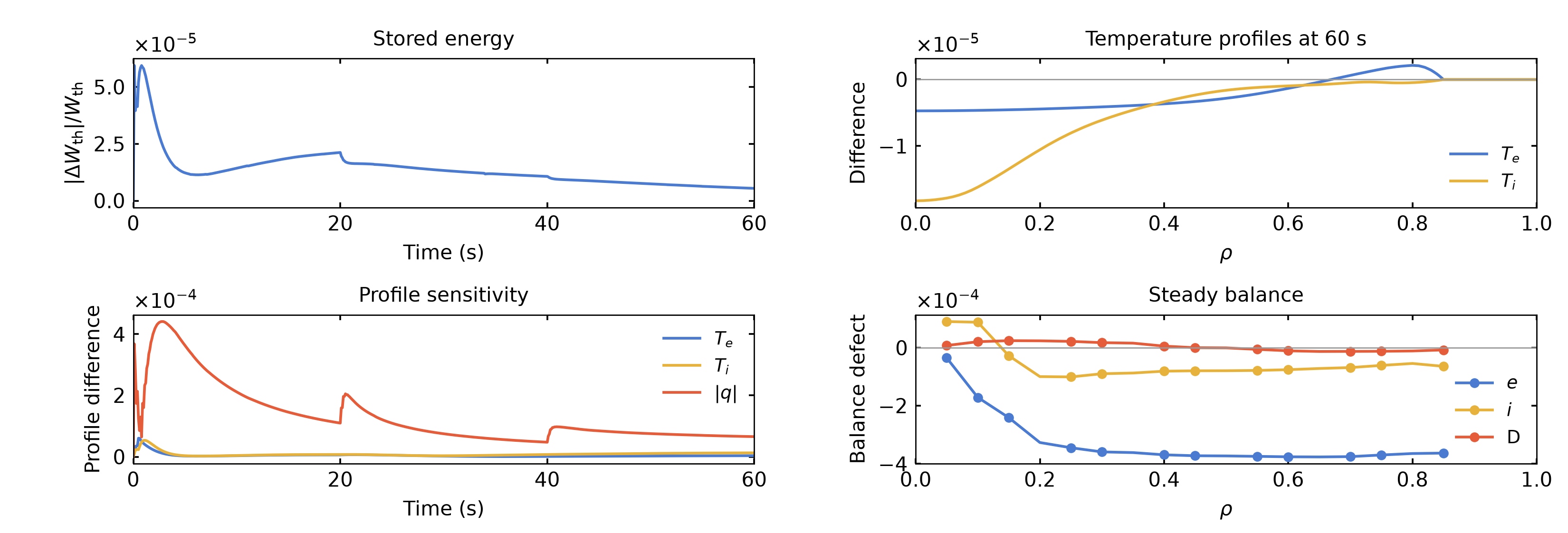}
  \caption{Numerical checks for the adaptive 60~s VTS pulse on 17 transport surfaces. The left column compares the 0.1~s coupling interval with the 0.05~s reference: the upper panel shows the relative difference in stored energy and the lower panel the symmetric profile differences of Eq.~\eqref{eq:profile-difference} for $T_e$, $T_i$ and $|q|$. The right column shows the relative $T_e$ and $T_i$ differences at 60~s against $\rho$ (upper) and the steady source--flux defect against $\rho$, normalized by the maximum absolute integrated source in each channel (lower). All quantities are dimensionless.}
  \label{fig:verification}
\end{figure}

Figure~\ref{fig:verification} assesses the temporal sensitivity and the steady balance defects. The largest steady source--flux defects, normalized to the maximum integrated source in each channel, are 0.038\% for electrons, 0.010\% for ions and 0.0025\% for particles.

Halving the coupling interval from 0.1 to 0.05~s, with adaptive TR-BDF2 in both subsystems and the same 17 transport surfaces, changes stored energy and the safety-factor profile by at most $5.95\times10^{-5}$ and $4.40\times10^{-4}$ over common sampled times. The largest electron- and ion-temperature profile differences are $6.09\times10^{-5}$ and $5.41\times10^{-5}$. These VTS changes are smaller than the corresponding VTS--FUSE differences at 60~s, so the tested VTS coupling interval alone does not explain the cross-code discrepancy. Spatial discretization, closure evaluation and current reconstruction remain possible contributors. \ref{app:time-integration} examines internal step selection and sensitivity to the thermal time tolerance. Radial convergence of the full pulse and FUSE time-step sensitivity remain to be assessed for this scenario.

An independent reconstruction checks the energy budget across the pulse. The net source power minus the outward thermal power at $\rho=0.85$ is integrated from the saved coupling-endpoint profiles by trapezoidal quadrature, with corrections for the intervening equilibrium updates and the prescribed outer region:
\begin{equation}
  \begin{aligned}
    W_{\rm bal}^{n+1} & =W_{\rm bal}^{n}+\frac{\Delta t}{2}
    \left[(P_{\rm src}-P_{\rm out})_n+(P_{\rm src}-P_{\rm out})_{n+1}\right]   \\
                      & \quad+\Delta W_{\rm geometry}+\Delta W_{\rm pedestal}.
  \end{aligned}
  \label{eq:energy-reconstruction}
\end{equation}
Starting from the initial thermal energy, with the first power balance evaluated after the heating transition, this gives $W_{\rm bal}=250.39$~MJ at 60~s against 252.70~MJ from direct volume integration, a difference of 0.91\%. The discrepancy is much larger than the stored-energy change under coupling-interval refinement: the latter tests the coupled evolution, while the reconstruction also depends on sampled power histories and separate geometry and pedestal corrections. The rapid initial heating makes power-history quadrature a plausible contributor, but its share cannot be separated without a finer diagnostic power history.

\subsection{Computational performance}
\label{sec:performance}

\begin{figure}[!htbp]
  \centering
  \includegraphics[width=1.00\textwidth]{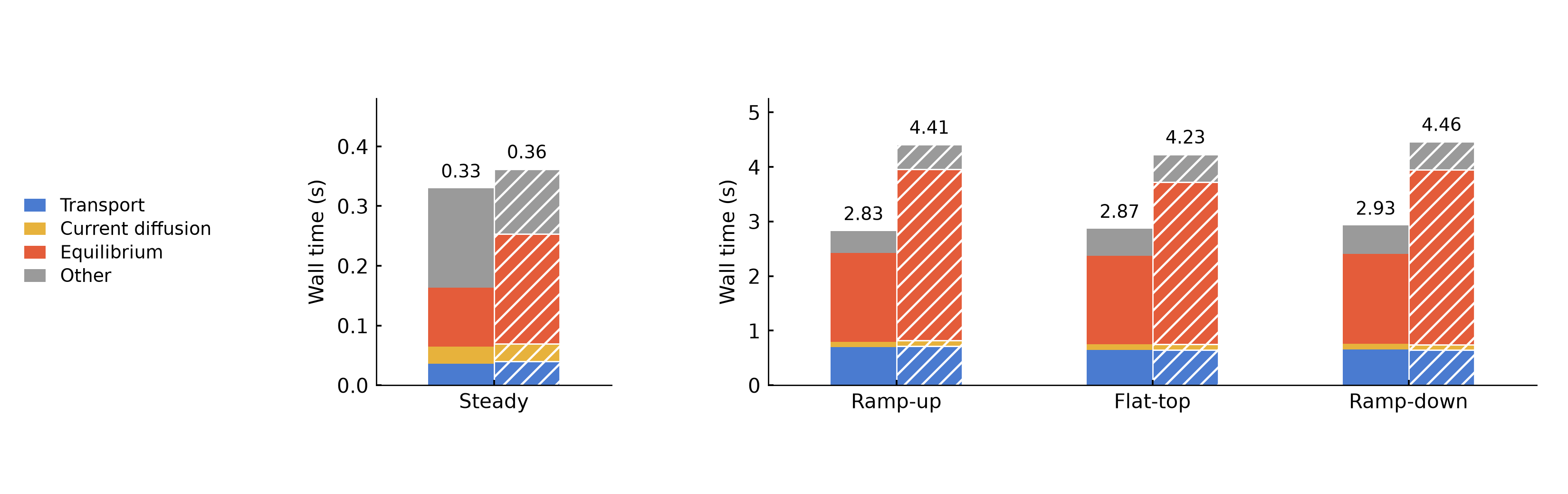}
  \caption{VTS wall time by phase. Each group compares the standard run with VTS (cold), which disables cross-interval solution reuse. The stacked bars show absolute time for transport, current diffusion, equilibrium, and other coupling work.}
  \label{fig:performance}
\end{figure}

The coupled pulse advances 600 coupling intervals of 0.1~s in 8.62~s of total wall time, corresponding to an average turnaround of approximately 14~ms per coupling interval (which encompasses two equilibrium updates, one transport advance, and one current-diffusion solve). This puts the single coupled step execution firmly on the 10~ms scale. Figure~\ref{fig:performance} and Table~\ref{tab:pulse-cost} resolve this cost by phase and subsystem. The equilibrium updates are the largest measured cost in each time-dependent phase.

Two key numerical factors enable this second-scale simulation capability:
\begin{enumerate}
    \item \textbf{Millisecond-scale variational equilibrium:} By exploiting the direct MXH--Chebyshev variational formulation and reusing preceding converged coefficients, the VEQ module solves each fixed-boundary equilibrium in approximately 5~ms. This is orders of magnitude faster than conventional grid-based finite-difference or Picard Grad--Shafranov solvers, which typically require hundreds of milliseconds.
    \item \textbf{Optimized one-dimensional transport formulation:} The transport subsystem is tailored for rapid execution by solving for logarithmic gradients on sparse flux surfaces, employing an analytically preconditioned Newton--Krylov (GMRES) solver, and advancing inventories via adaptive TR-BDF2 time stepping. This allows the stiff multi-channel transport update to complete in only $\sim$3~ms per advance.
\end{enumerate}
Together with the efficient current diffusion, the coupling overhead between equilibrium and transport is minimized, allowing the entire 60~s multi-timescale discharge to be simulated in a matter of seconds.
Reusing preceding converged solutions reduces the cost of successive solves. Disabling this reuse raises the pulse time from 8.62 to 13.10~s, principally through the equilibrium cost, which grows from 4.90 to 9.31~s; transport remains at 2.01~s. Both runs accept 612 transport substeps and reject seven trials, and their temperature profiles differ by at most $8.9\times10^{-8}$.

The FUSE calculation takes 790.99~s for the steady initialization and 2724.06~s for the pulse. Transport dominates its pulse cost at 1840.58~s (67.6\%), while equilibrium accounts for 867.85~s. Both calculations were prewarmed by a steady solve and one heated time step, and the timings were measured on an Apple M4 Pro with 48~GiB memory, excluding input preparation and output. Their different discretizations, tolerances and threading settings limit a direct attribution of the wall-time difference to any single method.
The time required for FUSE to reach a steady state is sensitive to the initial profiles. In tests of other cases with similar configurations, steady-state wall times ranged from approximately 100 to 1000~s. Transport solves accounted for most of the cost and its substantial variation. The steady-state timing in Fig.~\ref{fig:performance} is therefore specific to the initial conditions of this comparison.

\begin{table}[!htbp]
  \tableformat
  \caption{VTS cost by phase and subsystem. Each entry gives wall time in seconds followed by the percentage of the phase total. VTS (cold) disables cross-interval solution reuse; ``Other'' collects coupling and source updates.}
  \label{tab:pulse-cost}
  \begin{tabular}{llrrrr}
    \toprule
    Phase     & Subsystem         & \multicolumn{2}{c}{VTS} & \multicolumn{2}{c}{VTS (cold)}                         \\
    \cmidrule(lr){3-4}\cmidrule(lr){5-6}
              &                   & Time (s)                & Share (\%)                     & Time (s) & Share (\%) \\
    \midrule
    Steady    & Transport         & 0.036                   & 11.0                           & 0.040    & 11.1       \\
              & Current diffusion & 0.029                   & 8.8                            & 0.030    & 8.3        \\
              & Equilibrium       & 0.099                   & 29.9                           & 0.183    & 50.7       \\
              & Other             & 0.166                   & 50.3                           & 0.108    & 30.0       \\
    \addlinespace
    Ramp-up   & Transport         & 0.701                   & 24.8                           & 0.718    & 16.3       \\
              & Current diffusion & 0.097                   & 3.4                            & 0.105    & 2.4        \\
              & Equilibrium       & 1.622                   & 57.4                           & 3.133    & 71.1       \\
              & Other             & 0.407                   & 14.4                           & 0.452    & 10.3       \\
    \addlinespace
    Flat-top  & Transport         & 0.650                   & 22.7                           & 0.651    & 15.4       \\
              & Current diffusion & 0.099                   & 3.4                            & 0.100    & 2.4        \\
              & Equilibrium       & 1.625                   & 56.6                           & 2.973    & 70.3       \\
              & Other             & 0.495                   & 17.2                           & 0.505    & 11.9       \\
    \addlinespace
    Ramp-down & Transport         & 0.659                   & 22.5                           & 0.646    & 14.5       \\
              & Current diffusion & 0.099                   & 3.4                            & 0.098    & 2.2        \\
              & Equilibrium       & 1.649                   & 56.4                           & 3.200    & 71.8       \\
              & Other             & 0.518                   & 17.7                           & 0.516    & 11.6       \\
    \bottomrule
  \end{tabular}
\end{table}

\section{Discussion and conclusions}
\label{sec:discussion}

The heated pulse separates two aspects of the coupled evolution. Stored energy responds within a few seconds as increased outward transport balances the additional EC power. During the 20~s flat-top, the current enclosed by $\rho=0.85$ rises from 12.58 to 13.65~MA while total current remains fixed. The internal magnetic state thus continues to evolve after the thermal response has largely settled.
Recent FUSE releases provide a VEQ-based direct equilibrium solver as the default. The present comparison uses the Picard equilibrium solver of FUSE~1.1.5 to retain an independent equilibrium method alongside the VTS variational calculation.
Against the independent FUSE calculation, the stored energies differ by 0.02\% at 60~s, and the electron- and ion-temperature profiles by 0.37\% and 1.23\%. The toroidal-current profiles differ by 16.79\% over the full radius, with the largest differences near the axis. VTS coupling-interval refinement changes its profiles by less than this cross-code difference. The cumulative IMAS--QED current-transfer error limits the interpretation of the core-current comparison; further assessment requires consistent current transfer and radial refinement. The 0.91\% energy-reconstruction discrepancy also calls for a finer power history before assigning a global conservation accuracy.


The simulation of a 60~s dynamic discharge in 8.62~s demonstrates that coupled profile and equilibrium evolution can be achieved on a second-scale computational turnaround. This high efficiency is primarily underpinned by the millisecond-scale execution of the veloce equilibrium solver (VEQ), which is several orders of magnitude faster than standard Grad--Shafranov solvers, combined with an optimized one-dimensional transport scheme that achieves a single coupled advance on the order of 10~ms. 
Under this framework, the numerical coupling between magnetic geometry and core profiles is no longer the pacing computational bottleneck. Instead, as integrated modelling advances toward higher physical fidelity, the principal challenge shifts to the availability of {fast yet high-fidelity} physical closures. Standard nonlinear or even quasilinear gyrokinetic transport models and kinetic heating/current-drive codes (e.g., Fokker--Planck or Monte Carlo neutral beam modules) remain computationally prohibitive for second-scale multi-timescale simulations. Bridging this gap will require developing accelerated high-fidelity closures, that can be integrated into this rapid variational coupling architecture without eroding its second-scale efficiency.

\section*{Acknowledgements}

This research is supported by the National MCF Energy R\&D Program of China (Grant Nos. 2024YFF03000102 and 2022YFE03090000) and the National Natural Science Foundation of China (Grant No. 12435014).

We thank the developers of FUSE, QED.jl, and IMAS.jl for making their source codes openly accessible, which provided valuable reference and guidance during the development of this work. We are also grateful to Xiaohe Wu for insightful discussions.

\FloatBarrier
\appendix
\setcounter{figure}{0}
\renewcommand{\thefigure}{A\arabic{figure}}
\renewcommand{\theHfigure}{A.\arabic{figure}}

\section{Nonlinear transport solve}
\label{app:newton-krylov}

At fixed geometry, the particle and energy equations are solved for the logarithmic gradients $\boldsymbol z$ introduced in Eq.~\eqref{eq:reconstruction}. An implicit stage uses the conservative residual
\begin{equation}
  \boldsymbol F(\boldsymbol z)=D^{-1}\{h[C\boldsymbol s(\boldsymbol z)-\boldsymbol Q(\boldsymbol z)+\boldsymbol H]
  -a[\boldsymbol Y(\boldsymbol z)-\boldsymbol Y^n]\},
  \label{eq:nk-stage}
\end{equation}
where $\boldsymbol Q$ contains area-integrated outward fluxes, $\boldsymbol Y$ the cumulative particle and energy inventories, and $\boldsymbol H$ the known stage contribution; $h$ is the time step and $a$ is fixed by the time scheme. $D$ fixes one scale per channel from the initial cumulative target and throughput. The steady residual omits the inventory and history terms and the factor $h$.

Let $\boldsymbol\ell_s$ and $\boldsymbol\ell_f$ denote log profiles on the source and flux grids. Their perturbations follow from Eq.~\eqref{eq:reconstruction}:
\begin{equation}
  \delta\boldsymbol\ell_s=L_s\boldsymbol v,\qquad
  \delta\boldsymbol\ell_f=L_f\boldsymbol v,\qquad
  \delta\boldsymbol y=\operatorname{diag}(\boldsymbol y)L\boldsymbol v,\qquad
  \boldsymbol v=\delta\boldsymbol z .
  \label{eq:nk-reconstruction}
\end{equation}
With local closures, the Jacobian action is
\begin{equation}
  J\boldsymbol v=D^{-1}\{h[CS_\ell L_s\boldsymbol v-Q_z\boldsymbol v-Q_\ell L_f\boldsymbol v]
  -aY_z\boldsymbol v\},
  \label{eq:nk-product}
\end{equation}
where $S_\ell=\partial\boldsymbol s/\partial\boldsymbol\ell_s$, $Q_z=\partial\boldsymbol Q/\partial\boldsymbol z$ and $Q_\ell=\partial\boldsymbol Q/\partial\boldsymbol\ell_f$, all at fixed geometry. The inventory derivative $Y_z=\partial\boldsymbol Y/\partial\boldsymbol z$ includes quasineutrality and density--temperature coupling. External sources are fixed during a nonlinear solve.

The generalized minimal residual method (GMRES)\cite{saad1986gmres} solves $J\boldsymbol p=-\boldsymbol F$ with a preconditioner that retains the integrated source and inventory response and the same-surface flux blocks,
\begin{equation}
  B=D^{-1}(hCS_\ell L_s-aY_z),\qquad
  M=B-hD^{-1}\mathcal D_f(Q_z+Q_\ell L_f),
  \label{eq:nk-preconditioner}
\end{equation}
in which $\mathcal D_f$ selects same-surface blocks. The linear tolerance is $\|J\boldsymbol p+\boldsymbol F\|_2\leq10^{-10}\|\boldsymbol F\|_2$. Forward differences perturb gradients by $10^{-6}\max(1,|z_{c,j}|)$ and log profiles by $10^{-6}$. Logarithmic reconstruction preserves positive densities and temperatures. Backtracking accepts a trial only when the closure remains valid and
\begin{equation}
  \|\boldsymbol F(\boldsymbol z+\alpha\boldsymbol p)\|_2
  <(1-10^{-4}\alpha)\|\boldsymbol F(\boldsymbol z)\|_2 .
  \label{eq:nk-linesearch}
\end{equation}
For $N$ unknowns, convergence requires $\|\boldsymbol F\|_2/\sqrt N\leq10^{-6}+10^{-8}$ and $\|\boldsymbol p/(1+|\boldsymbol z|)\|_2/\sqrt N\leq\tau_z$, where $\tau_z=10^{-6}$ for steady solves and $10^{-12}$ for implicit stages. An exactly zero residual terminates directly. The line search respects $|z_i|\leq30$, starting within 99.5\% of the distance to these bounds. Small inventory changes are evaluated from profile increments using $\Delta y=y^n\operatorname{expm1}(\Delta\ln y)$, where $\operatorname{expm1}(x)=\exp(x)-1$ is evaluated without cancellation near zero. Failed stages trigger step reduction.

\setcounter{figure}{0}
\renewcommand{\thefigure}{B\arabic{figure}}
\renewcommand{\theHfigure}{B.\arabic{figure}}

\section{Adaptive implicit integration}
\label{app:time-integration}

Both transport and current diffusion use TR-BDF2. For an inventory or current mass coordinate $y$ with rate $R$, the method evaluates $R$ at $t_n$, $t_n+ch$ and $t_n+h$, with $\gamma=1-1/\sqrt{2}$ and $c=2\gamma$. The update is $y^{n+1}=y^n+h\sum_{j=0}^2b_jR_j$, where $b=((1-\gamma)/2,(1-\gamma)/2,\gamma)$. The embedded weights are $\widehat b_1=[6c(1-c)]^{-1}$, $\widehat b_2=(2-3c)/[6(1-c)]$ and $\widehat b_0=1-\widehat b_1-\widehat b_2$. They give a defect $d=h\sum_j(b_j-\widehat b_j)R_j$ without another solve\cite{hosea1996}. A step is accepted when the implicit stages converge and
\begin{equation}
  E=\max_{k,i}\frac{|d_{k,i}|}
  {\epsilon_a\max_j|y^n_{k,j}|+\epsilon_r\max(|y^n_{k,i}|,|y^{n+1}_{k,i}|)}
  \leq1,
  \label{eq:time-error}
\end{equation}
with $\epsilon_r=10^{-4}$ and $\epsilon_a=10^{-8}$. The next step follows $0.9E^{-1/3}$, bounded by factors $(0.5,2)$ for transport and $(0.1,4)$ for current diffusion. Initial steps are 0.1~s and $10^{-3}$~s, with regular lower bounds of $10^{-4}$ and $10^{-6}$~s, respectively. The coupling interval supplies the upper bound; rejected steps shrink by at least a factor of two. Current diffusion applies Eq.~\eqref{eq:time-error} to $M\boldsymbol w$, excluding the constrained boundary rows.

The adaptive control is exercised almost entirely by the heating transition. The first candidate step of the ramp-up interval gives $E=193$ and is resolved by eight accepted substeps, and all seven rejected transport trials in the pulse occur before 0.4~s, each with converged stage solves rather than a solver failure. Over the remaining 596 of the 600 intervals a single 0.1~s step is accepted, so the refinement is confined to where the source changes. In total, transport accepts 612 substeps and current diffusion 3109, the latter without any rejections, and the smallest accepted steps are $2.17\times10^{-3}$ and $1.08\times10^{-4}$~s. Tightening the thermal tolerance from $10^{-4}$ to $10^{-5}$ changes the electron-temperature profile by at most $5.10\times10^{-5}$ and the stored energy by $3.02\times10^{-5}$, so the default tolerance is not limiting the thermal solution.

\begin{figure}[!htbp]
  \centering
  \includegraphics[width=1.00\textwidth]{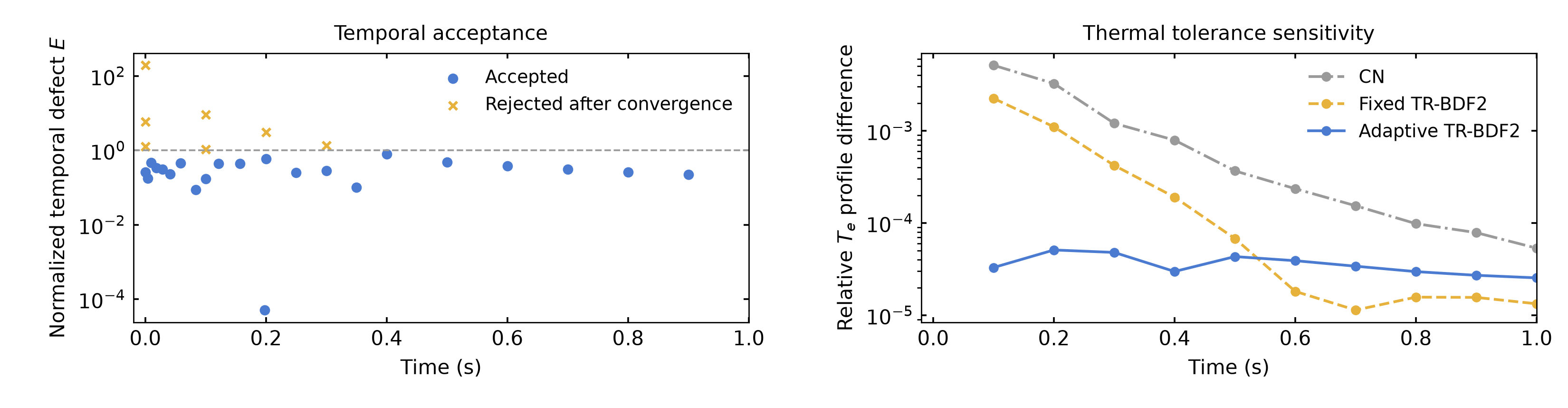}
  \caption{Adaptive integration through the heating transition. Left: the normalized transport error estimate $E$ of Eq.~\eqref{eq:time-error} for the accepted (blue) and rejected (yellow) trial steps; all stage solves converged, and the dashed line marks $E=1$. Right: symmetric electron-temperature differences from the calculation with thermal tolerance $10^{-5}$, evaluated with Eq.~\eqref{eq:profile-difference}, for Crank--Nicolson and fixed-step TR-BDF2 against adaptive TR-BDF2. The current-diffusion tolerance and the 0.1~s coupling interval are unchanged throughout.}
  \label{fig:adaptive-control}
\end{figure}


\begin{thebibliography}{99}
  \bibitem{bourdelle2025} C. Bourdelle, Integrated modelling of tokamak plasmas: progress and challenges towards ITER operation and reactor design, Plasma Physics and Controlled Fusion 67 (2025) 043001. \url{https://doi.org/10.1088/1361-6587/adc484}.
  \bibitem{artaud2010cronos} J. F. Artaud et al., The CRONOS suite of codes for integrated tokamak modelling, Nuclear Fusion 50 (2010) 043001. \url{https://doi.org/10.1088/0029-5515/50/4/043001}.
  \bibitem{fable2013astra} E. Fable et al., Novel free-boundary equilibrium and transport solver with theory-based models and its validation against ASDEX Upgrade current ramp scenarios, Plasma Physics and Controlled Fusion 55 (2013) 124028. \url{https://doi.org/10.1088/0741-3335/55/12/124028}.
  \bibitem{lyons2023step} B. C. Lyons et al., Flexible, integrated modeling of tokamak stability, transport, equilibrium, and pedestal physics, Physics of Plasmas 30 (2023) 092510. \url{https://doi.org/10.1063/5.0156877}.
  \bibitem{pankin2025transp} A. Y. Pankin et al., TRANSP integrated modeling code for interpretive and predictive analysis of tokamak plasmas, Computer Physics Communications 312 (2025) 109611. \url{https://doi.org/10.1016/j.cpc.2025.109611}.
  \bibitem{pankin2026predictive} A. Y. Pankin et al., Predictive capabilities of the integrated modeling TRANSP code for tokamak plasmas, Computer Physics Communications 328 (2026) 110337. \url{https://doi.org/10.1016/j.cpc.2026.110337}.
  \bibitem{imbeaux2015imas} F. Imbeaux et al., Design and first applications of the ITER integrated modelling \& analysis suite, Nuclear Fusion 55 (2015) 123006. \url{https://doi.org/10.1088/0029-5515/55/12/123006}.
  \bibitem{barnes2010trinity} M. Barnes et al., Direct multiscale coupling of a transport code to gyrokinetic turbulence codes, Physics of Plasmas 17 (2010) 056109. \url{https://doi.org/10.1063/1.3323082}.
  \bibitem{disiena2022genetango} A. Di Siena et al., Global gyrokinetic simulations of ASDEX Upgrade up to the transport timescale with GENE--Tango, Nuclear Fusion 62 (2022) 106025. \url{https://doi.org/10.1088/1741-4326/ac8941}.
  \bibitem{fastran2017} J. M. Park et al., An efficient transport solver for tokamak plasmas, Computer Physics Communications 214 (2017) 1--5. \url{https://doi.org/10.1016/j.cpc.2016.12.018}.
  \bibitem{felici2011raptor} F. Felici et al., Real-time physics-model-based simulation of the current density profile in tokamak plasmas, Nuclear Fusion 51 (2011) 083052. \url{https://doi.org/10.1088/0029-5515/51/8/083052}.
  \bibitem{felici2012raptor} F. Felici et al., Non-linear model-based optimization of actuator trajectories for tokamak plasma profile control, Plasma Physics and Controlled Fusion 54 (2012) 025002. \url{https://doi.org/10.1088/0741-3335/54/2/025002}.
  \bibitem{citrin2024torax} J. Citrin et al., TORAX: A Fast and Differentiable Tokamak Transport Simulator in JAX, arXiv preprint (2024). \url{https://doi.org/10.48550/arXiv.2406.06718}.
  \bibitem{meneghini2024fuse} O. Meneghini et al., FUSE (Fusion Synthesis Engine): A Next Generation Framework for Integrated Design of Fusion Pilot Plants, arXiv preprint (2024). \url{https://doi.org/10.48550/arXiv.2409.05894}.
  \bibitem{artaud2018metis} J. F. Artaud et al., Metis: a fast integrated tokamak modelling tool for scenario design, Nuclear Fusion 58 (2018) 105001. \url{https://doi.org/10.1088/1741-4326/aad5b1}.
  \bibitem{xie2026mxh} H. Xie and Y. Li, Compact MXH-Chebyshev representation of fixed-boundary tokamak equilibria, Plasma Physics and Controlled Fusion 68 (2026) 085021. \url{https://doi.org/10.1088/1361-6587/ae9591}.
  \bibitem{zhang2026veq} R. Zhang et al., VEQ: a fast parametric Grad--Shafranov solver for fixed-boundary tokamak equilibria with flexible source profiles, arXiv preprint (2026). \url{https://doi.org/10.48550/arXiv.2606.11821}.
  \bibitem{erba1998} M. Erba et al., Validation of a new mixed Bohm/gyro-Bohm model for electron and ion heat transport against the ITER, Tore Supra and START database discharges, Nuclear Fusion 38 (1998) 1013--1028. \url{https://doi.org/10.1088/0029-5515/38/7/305}.
  \bibitem{chang1982} C. S. Chang and F. L. Hinton, Effect of finite aspect ratio on the neoclassical ion thermal conductivity in the banana regime, The Physics of Fluids 25 (1982) 1493--1494. \url{https://doi.org/10.1063/1.863934}.
  \bibitem{redl2021} A. Redl et al., A new set of analytical formulae for the computation of the bootstrap current and the neoclassical conductivity in tokamaks, Physics of Plasmas 28 (2021) 022502. \url{https://doi.org/10.1063/5.0012664}.
  \bibitem{haney1992supercode} S. W. Haney et al., A ``SuperCode'' for Systems Analysis of Tokamak Experiments and Reactors, Fusion Technology 21 (1992) 1749--1758. \url{https://doi.org/10.13182/FST92-A29974}.
  \bibitem{candy2009tgyro} J. Candy, C. Holland, R. E. Waltz, M. R. Fahey and E. Belli, Tokamak profile prediction using direct gyrokinetic and neoclassical simulation, Physics of Plasmas 16 (2009) 060704. \url{https://doi.org/10.1063/1.3167820}.
  \bibitem{jardin2010} S. Jardin, Computational Methods in Plasma Physics, CRC Press, 2010. \url{https://doi.org/10.1201/EBK1439810958}.
  \bibitem{bank1985} R. E. Bank et al., Transient simulation of silicon devices and circuits, IEEE Transactions on Computer-Aided Design of Integrated Circuits and Systems 4 (1985) 436--451. \url{https://doi.org/10.1109/TCAD.1985.1270142}.
  \bibitem{hosea1996} M. E. Hosea and L. F. Shampine, Analysis and implementation of TR-BDF2, Applied Numerical Mathematics 20 (1996) 21--37. \url{https://doi.org/10.1016/0168-9274(95)00115-8}.
  \bibitem{sauter2013cocos} O. Sauter and S. Yu. Medvedev, Tokamak coordinate conventions: COCOS, Computer Physics Communications 184 (2013) 293--302. \url{https://doi.org/10.1016/j.cpc.2012.09.010}.
  \bibitem{saad1986gmres} Y. Saad and M. H. Schultz, GMRES: A generalized minimal residual algorithm for solving nonsymmetric linear systems, SIAM Journal on Scientific and Statistical Computing 7 (1986) 856--869. \url{https://doi.org/10.1137/0907058}.
\end{thebibliography}
\end{document}